\documentclass[envcountsame]{llncs}
\usepackage{amsmath}
\usepackage{amssymb}
\usepackage{tikz}
\usepackage{color}
\usepackage{gnuplot-lua-tikz}
\usepackage{url}

\let\set\mathbb
\def\<#1>{\langle#1\rangle}

\def\eatspace#1{#1}
\def\step#1#2{\par\kern1pt\dimen44=#2em\advance\dimen44 1.67em\hangindent\dimen44\hangafter=1\noindent\rlap{\small#1}\kern\dimen44\relax\eatspace}

\begin{document}

 \author{Marijn J.H. Heule\inst{1}\thanks{Supported by NSF grant CCF-1813993 and AFRL Award FA8750-15-2-0096.} \and
Manuel Kauers\inst{2}\thanks{Supported by the Austrian FWF grants P31571-N32 and F5004.} \and
Martina Seidl\inst{3}\thanks{Supported by the Austrian FWF grant NFN S11408-N23 
and the LIT AI Lab funded by the State of Upper Austria.}}

\institute{Department of Computer Science, The University of Texas at Austin, United States\and
Institute for Algebra, J. Kepler University Linz, Austria\and
Institute for Formal Models and Verification, J. Kepler University Linz, Austria}

 \title{Local Search for Fast Matrix Multiplication}

 \maketitle

  \begin{abstract}
Laderman discovered a scheme for computing the product of two $3\times3$ matrices
using only 23 multiplications in 1976. Since then, some
more such schemes were proposed, but nobody knows how many such schemes
there are and whether there exist schemes with fewer than 23
multiplications. In this paper we present two independent
SAT-based methods for finding new schemes using 23 multiplications. 
Both methods allow
computing a few hundred new schemes individually, and many
thousands when combined. Local search SAT solvers outperform CDCL
solvers consistently in this application.
 \end{abstract}

\section{Introduction}

Matrix multiplication is a fundamental operation with applications in nearly any
area of science and engineering. However, after more than 50 years of work on
matrix multiplication techniques (see,
e.g.,~\cite{buergisser-book,pan-survey2018,blaeser-survey2013,landsberg2017geometry}),
the complexity of matrix multiplication is still a mystery. Even for small matrices, the problem
is not completely understood, and understanding these cases better can provide valuable hints
towards more efficient algorithms for large matrices.

The naive way for computing the product $C$ of two $2\times 2$ matrices
$A,B$ requires 8 multiplications:
\[
 \begin{pmatrix}
 a_{11} & a_{12} \\
 a_{21} & a_{22}
 \end{pmatrix}
 \begin{pmatrix}
 b_{11} & b_{12} \\
 b_{21} & b_{22}
 \end{pmatrix} =
 \begin{pmatrix}
 a_{11}b_{11} + a_{12}b_{21} &\kern.5em a_{11}b_{12} + a_{21}b_{22}\\
 a_{21}b_{11} + a_{22}b_{21} &\kern.5em a_{21}b_{12} + a_{22}b_{22}
 \end{pmatrix}
 =
 \begin{pmatrix}
 c_{11} & c_{12} \\
 c_{21} & c_{22}
 \end{pmatrix}
\]
Strassen observed 50 years ago that $C$ can also be computed with only 7
multiplications~\cite{strassen1969gaussian}.
His scheme proceeds in two steps.
In the first step he introduces auxiliary variables $M_1, \ldots, M_7$
which are defined as the product of certain linear combinations of
the entries of $A$ and~$B$. In the second step the entries of $C$ are obtained as
certain linear combinations of the~$M_i$:
\begin{alignat*}{5}
 M_1 &= (a_{11} + a_{22}) (b_{11} + b_{22})
 \qquad\qquad & c_{11} & =  M_1 + M_4 - M_5 + M_7 \\[-2pt]
 M_2 &= (a_{21} + a_{22}) (b_{11})
  & c_{12} & =  M_3 + M_5 \\[-2pt]
 M_3 &= (a_{11}) (b_{12} - b_{22})
  & c_{21} & =   M_2 + M_4 \\[-2pt]
 M_4 &= (a_{22}) (b_{21} - b_{11})
  & c_{22} & =   M_1 - M_2 + M_3 + M_6 \\[-2pt]
 M_5 &= (a_{11} + a_{12})(b_{22})\\[-2pt]
 M_6 &= (a_{21} - a_{11}) (b_{11}+ b_{12})\\[-2pt]
 M_7 &= (a_{12} - a_{22}) (b_{21} + b_{22})
\end{alignat*}
Recursive application of this scheme gave rise to the first algorithm for multiplying
arbitrary $n\times n$ matrices in subcubic complexity.
Winograd~\cite{winograd1971multiplication} showed
that Strassen's scheme is optimal in the sense
that there does not exist a similar scheme with fewer than 7 multiplications,
and de Groote~\cite{de1978varieties}
showed that Strassen's scheme is essentially unique.

Less is known about $3\times 3$ matrices.
The naive scheme requires 27 multiplications, and in 1976
Laderman~\cite{laderman1976noncommutative} found one with~23.
Similar as Strassen, he defines $M_1,\dots,M_{23}$ as
products of certain linear combinations of
the entries of $A$ and~$B$. The entries of $C=AB$ are then obtained
as linear combinations of $M_1,\dots,M_{23}$.
It is not known whether 23 is optimal (the best lower bound is 19~\cite{blaser2003complexity}).
It is known however that Laderman's scheme is \emph{not} unique.
A small number of intrinsically different schemes have been found
over the years. Of particular interest are schemes in which all
coefficients in the linear combinations are $+1$, $-1$, or~$0$.
The only four such schemes (up to equivalence) we are aware of
are due to Laderman, Smirnov~\cite{smirnov2013bilinear}, Oh et al.~\cite{oh2013inequivalence}, and Courtois
et al.~\cite{DBLP:journals/corr/abs-1108-2830}. 

While Smirnov and Oh et al. found their multiplicatoin schemes 
with computer-based search using non-linear numerical optimization 
methods, Courtois found his multiplication scheme using a SAT solver. This is
also what we do here. We present two approaches which allowed us to generate more
than 13,000 mutually inequivalent new matrix multiplication schemes for $3\times
3$ matrices, using altogether about 35 years of CPU years. We believe that the
new schemes are of interest to the matrix multiplication community. We therefore
make them publicly available in various formats and grouped by invariants at
\begin{center}
\url{http://www.algebra.uni-linz.ac.at/research/matrix-multiplication/}.
\end{center}

\section{Encoding and Workflow}

\def\x{^{\vphantom(}} 

To search for multiplication schemes of $3 \times 3$ matrices having
the above form, we define the $M_i$ as product of linear combination
of all entries of $A$ and $B$ with undetermined coefficients $\alpha_{ij}^{(\ell)}, \beta_{ij}^{(\ell)}$:
 \begin{alignat*}1
   M_1\x &=
   (\alpha_{11}^{(1)}a_{11}\x + \cdots + \alpha_{33}^{(1)}a_{33}\x)
   (\beta_{11}^{(1)}b_{11}\x + \cdots + \beta_{33}^{(1)}b_{33}\x)\\[-2pt]
   &\vdots\\[-2pt]
   M_{23}\x &=
   (\alpha_{11}^{(23)}a_{11}\x + \cdots + \alpha_{33}^{(23)}a_{33}\x)
   (\beta_{11}^{(23)}b_{11}\x + \cdots + \beta_{33}^{(23)}b_{33}\x)
 \end{alignat*}
 Similarly, we define the $c_{ij}$ as linear combinations of the
 $M_i$ with undetermined coefficients $\gamma_{i,j}^{(\ell)}$:
 \begin{alignat*}3
   c_{11}\x &= \gamma_{11}^{(1)}M_1\x + \cdots + \gamma_{11}^{(23)}M_{23}\x,
   &\quad \ldots,\quad  &
   c_{33}\x &= \gamma_{33}^{(1)}M_1\x + \cdots + \gamma_{33}^{(23)}M_{23}\x
 \end{alignat*}
 Comparing the coefficients of all terms $a_{i_1i_2}b_{j_1j_2}c_{k_1k_2}$ in the equations $c_{ij}=\sum_k a_{ik}b_{kj}$
 leads to the polynomial equations
 \[
 \sum_{\ell=1}^{23} \alpha_{i_1i_2}^{(\ell)}\beta_{j_1j_2}^{(\ell)}\gamma_{k_1k_2}^{(\ell)} =
 \delta_{i_2j_1}\x\delta_{i_1k_1}\x\delta_{j_2k_2}\x
 \]
 for $i_1,i_2,j_1,j_2,k_1,k_2\in\{1,2,3\}$.  These 729 cubic equations with 621
 variables are also known as \emph{Brent equations}~\cite{brent1970algorithms}.  The
 $\delta_{uv}$ on the right are Kronecker-deltas, i.e., $\delta_{uv}=1$ if $u=v$
 and $\delta_{uv}=0$ otherwise.  Each solution of the system of these equations
 corresponds to a matrix multiplication scheme.  The equations become slightly
 more symmetric if we flip the indices of the~$\gamma_{ij}$, and since this is
 the variant mostly used in the literature, we will also adopt it from now on.

 Another view on the Brent equations is as follows. View the
 $\alpha_{i_1i_2}^{(\ell)}$, $\beta_{j_1j_2}^{(\ell)}$,
 $\gamma_{k_1k_2}^{(\ell)}$ as variables, as before, and regard
 $a_{i_1i_2},b_{j_1j_2},c_{k_1k_2}$ as polynomial indeterminants. Then the task
 consists of instantiating the variables in such a way that
 \[
 \sum_{\ell=1}^{23}(\alpha_{11}^{(\ell)}a_{11}\x+\cdots)(\beta_{11}^{(\ell)}b_{11}\x+\cdots)(\gamma_{11}^{(\ell)}c_{11}\x+\cdots)
 =\sum_{i=1}^3\sum_{j=1}^3\sum_{k=1}^3 a_{ij}b_{jk}c_{ki}
 \]
 holds as equation of polynomials in the variables
 $a_{i_1i_2},b_{j_1j_2},c_{k_1k_2}$. Expanding the left hand side and equating
 coefficients leads to the Brent equations as stated before (but with indices of
 $\gamma$ flipped, as agreed). In other words, expanding the left hand side,
 all terms have to cancel except for the terms on the right. We found it
 convenient to say that a term $a_{i_1i_2}b_{j_1j_2}c_{k_1k_2}$ has ``type $m$''
 if $m = \delta_{i_2j_1}+\delta_{j_2k_1}+\delta_{k_2i_1}$.  With this
 terminology, all terms of types 0, 1,~2 have to cancel each other, and all
 terms of type~3 have to survive. Note that since all 27 type~3 terms
 must be produced by the 23 summands on the left, some summands must produce
 more than one type~3 term.

 For solving the Brent equations with a SAT solver, we use $\set Z_2$ as
 coefficient domain, so that multiplication translates into `and' and
 addition translates into `xor'. When, for example, the variable $\alpha_{i_1i_2}^{(\ell)}$ is true
 in a solution of the corresponding SAT instance, this indicates that
 the term $a_{i_1i_2}$ occurs in~$M_\ell$, and likewise for the $b$-variables.
 If $\gamma_{k_1k_2}^{(\ell)}$ is true, this means that $M_\ell$ appears in the
 linear combination for~$c_{k_1k_2}$. We call  $\alpha_{i_1i_2}^{(\ell)}$, $\beta_{j_1j_2}^{(\ell)}$, and $\gamma_{k_1k_2}^{(\ell)}$
 the \emph{base variables}.

 In order to bring the Brent equations into CNF, we use Tseitin transformation, i.e., we introduce definitions for
 subformulas to avoid exponential blow-up. To keep the number of fresh variables
 low, we do not introduce one new variable for every cube but only for pairs of literals,
 i.e., we encode a cube $(\alpha \land \beta \land \gamma)$ as $u \leftrightarrow (\alpha \land \beta)$ and
 $v \leftrightarrow (u \land \gamma)$. In this way, we can reuse~$u$.
 Furthermore, a sum $v_1 \oplus \dots \oplus v_m$ with $m\geq4$ is encoded by
 $w\leftrightarrow(v_1\oplus v_2\oplus v_3)$ and $v_4 \oplus \dots \oplus v_m\oplus w$,
 with the latter sum being encoded recursively. This encoding seems to require the
 smallest sum of the number of variables and the number of clauses---a commonly used optimality heuristic. The used scripts are available at 
\begin{center}
\url{https://github.com/marijnheule/matrix-challenges/tree/master/src}.
\end{center}
 The generation of new schemes proceeds in several steps, with SAT solving being
 the first and main step. If the SAT solver finds a solution, we next check
 whether it is equivalent to any known or previously found solution modulo de Groote's symmetry group~\cite{de1978varieties}.
 If so, we discard it. Otherwise, we next try to simplify the new scheme by searching for an element in
 its orbit which has a smaller number of terms. The scheme can then be used to initiate
 a new search. In the fourth step, we use
 Gr\"obner bases to lift the scheme from the coefficient domain $\set Z_2$ to~$\set Z$.
 Finally, we cluster large sets of similar solutions into parameterized families.
 
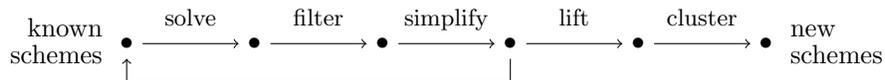
\begin{center}
   \begin{tikzpicture}[xscale=1.7]
     \node (b) at (1,0) {$\bullet$} (1,0) node[left,xshift=-2mm] {\vbox{\llap{\strut known}\kern-2pt\llap{\strut schemes}}};
     \node (c) at (2,0) {$\bullet$};
     \node (d) at (3,0) {$\bullet$};
     \node (e) at (4,0) {$\bullet$};
     \node (f) at (5,0) {$\bullet$};
     \node (g) at (6,0) {$\bullet$} (6,0) node[right,xshift=2mm] {\vbox{\rlap{\strut new}\kern-2pt\rlap{\strut schemes}}};
     \draw[->] (b)--(c) node[midway,above] {\footnotesize\strut solve};
     \draw[->] (c)--(d) node[midway,above] {\footnotesize\strut filter};
     \draw[->] (d)--(e) node[midway,above] {\footnotesize\strut simplify};
     \draw[->] (e)--(f) node[midway,above] {\footnotesize\strut lift};
     \draw[->] (f)--(g) node[midway,above] {\footnotesize\strut cluster};
     \draw[->] (e)--++(0,-.5)--++(-3,0)--(b);
   \end{tikzpicture}
 \end{center}
 In the present paper, we give a detailed description of the first step in this workflow.
 The subsequent steps use algebraic techniques unrelated to SAT and will be described in~\cite{mmr-jsc}.

 \section{Solving Methods}\label{sec:methods}

 The \emph{core} of a scheme is the pairing of the type~3 terms.
 Our first method focuses on finding schemes with new cores, while our second method searches for schemes that are similar to an
 existing one and generally has the same core.
 For all experiments we used the local search SAT solver {\tt yalsat}~\cite{yalsat} as this solver performed best on instances from this application.
 We also tried solving these instances using CDCL solvers, but the performance was disappointing. We observed a possible explanation:
 The runtime of CDCL solvers tends to be exponential in the average backtrack level (ABL) on unsatisfiable instances.
 For most formulas arising from other applications, ABL is small ($<50$), while on the matrix multiplication instances ABL is large ($> 100$).

 \subsection{Random Pairings of Type 3 Terms and Streamlining}

 Two of the known schemes, those of Smirnov~\cite{smirnov2013bilinear} and Courtois et al.~\cite{DBLP:journals/corr/abs-1108-2830},
 have the property that each type~3 term occurs exactly once and at most two type~3 terms occur in the same summand. We decided to use this pattern
  to search for new schemes: randomly pair four type~3 terms and assign the remaining type~3 terms to the other 19 summands.
  Only in very rare cases, random pairing could be extended to a valid scheme in reasonable time, say a few minutes.
  In the other cases it is not known whether the pairing cannot be extended to a valid scheme or whether finding such a scheme is very hard.

  Since the number of random pairings that could be extended to a valid scheme was very low, we
  tried 
  adding \emph{streamlining constraints}~\cite{streamlining} to formulas. A streamlining constraint is a set of clauses that
  guides the solver to a solution, but these clauses may not (and generally are not) implied by the formula. Streamlining constraints are
  usually patterns observed in solutions of a given problem, potentially of  smaller sizes. We experimented with various
  streamlining constraints, such as enforcing that each type~0, type~1, and type~2 term occurs either zero times or twice in a scheme
  (instead of an even number of times).  The most effective streamlining constraint that we came up with was observed in the Smirnov scheme:
  for each summand that is assigned a single type~3 term, enforce that (i) one matrix has either two rows, two columns
  or a row and a column fully assigned to zero and (ii) another matrix has two rows and two columns assigned to zero,
  i.e., the matrix has a single nonzero entry. This streamlining constraint reduced the runtime from minutes to seconds. Yet
  some random pairings may only be extended to a valid scheme that does not satisfy the streamlining constraint.

\subsection{Neighborhood Search}

The second method is based on neighborhood search: we select a scheme, randomly fix some the corresponding base variables,
and search for an assignment for the remaining base variables. This simple method turned out to be remarkably effective to find
new schemes. The only parameter for this method is the number of base variables that will be fixed. The lower the number of
fixed base variables, the higher the probability to find a different scheme and the higher the costs to find an assignment for the remaining base variables.
We experimented with various values and it turned out that fixing 2/3 of the 621 base variables (414) is effective to
find many new schemes in a reasonable time.

The neighborhood search is able to find many new schemes, but in almost all cases they have the same pairing
of type~3 terms. Only in some rare cases the pairing of type~3 terms is different. Figure~\ref{fig:neighbors}
shows such an example: scheme {\tt A} has term $a_{13}b_{31}c_{11}$ in summand 22 and
term $a_{23}b_{33}c_{32}$ in summand 23, while the neighboring scheme {\tt B} has term $a_{13}b_{33}c_{31}$ in summand 22
and terms $a_{13}b_{31}c_{11}$, $a_{23}b_{33}c_{32}$, and $a_{13}b_{33}c_{31}$ in summand~23.

\begin{figure}[ht]
\centering
\begin{minipage}{0.7\textwidth}\footnotesize
\begin{eqnarray*}
{\tt 1~\,}&~&(a_{11} + a_{13} + a_{21} + a_{22} + a_{23})(b_{13})(c_{22} + c_{32})\\[-2pt]
{\tt 2~\,}&~&(a_{11} + a_{13} + a_{23})(b_{13} + b_{32})(c_{11} + c_{22} + c_{31} + c_{32})\\[-2pt]
{\tt 3~\,}&~&(a_{11} + a_{13})(b_{32})(c_{21} + c_{22} + c_{31} + c_{32})\\[-2pt]
{\tt 4~\,}&~&(a_{11} + a_{31})(b_{11} + b_{12} + b_{13})(c_{23})\\[-2pt]
{\tt 5~\,}&~&(a_{11} + a_{33})(b_{11} + b_{13} + b_{32})(c_{11} + c_{23})\\[-2pt]
{\tt 6~\,}&~&(a_{12} + a_{13} + a_{23})(b_{13} + b_{33})(c_{11} + c_{31})\\[-2pt]
{\tt 7~\,}&~&(a_{12} + a_{22} + a_{32})(b_{21} + b_{22} + b_{23})(c_{33})\\[-2pt]
{\tt 8~\,}&~&(a_{12} + a_{31} + a_{32} + a_{33})(b_{22})(c_{23} + c_{33})\\[-2pt]
{\tt 9~\,}&~&(a_{12} + a_{33})(b_{13} + b_{21} + b_{33})(c_{11} + c_{33})\\[-2pt]
{\tt 10~\,}&~&(a_{12})(b_{13} + b_{23} + b_{33})(c_{31} + c_{33})\\[-2pt]
{\tt 11~\,}&~&(a_{21} + a_{31} + a_{33})(b_{11})(c_{12} + c_{22})\\[-2pt]
{\tt 12~\,}&~&(a_{21})(b_{11} + b_{12} + b_{13})(c_{22})\\[-2pt]
{\tt 13~\,}&~&(a_{22} + a_{31} + a_{33})(b_{13} + b_{22})(c_{12} + c_{13} + c_{22} + c_{33})\\[-2pt]
{\tt 14~\,}&~&(a_{22} + a_{32} + a_{33})(b_{21})(c_{13} + c_{33})\\[-2pt]
{\tt 15~\,}&~&(a_{22})(b_{13} + b_{21} + b_{22})(c_{12} + c_{13})\\[-2pt]
{\tt 16~\,}&~&(a_{22})(b_{13} + b_{23})(c_{32} + c_{33})\\[-2pt]
{\tt 17~\,}&~&(a_{23})(b_{31})(c_{11} + c_{12} + c_{31} + c_{32})\\[-2pt]
{\tt 18~\,}&~&(a_{31} + a_{33})(b_{11} + b_{13} + b_{22})(c_{12} + c_{13} + c_{22} + c_{23})\\[-2pt]
{\tt 19~\,}&~&(a_{33})(b_{11} + b_{21} + b_{31})(c_{11} + c_{13})\\ \hline
{\tt 20{A}}&~&(a_{12})(b_{22})(c_{21} + c_{23})\\[-2pt]
{\tt 21{A}}&~&(a_{11})(b_{12} + b_{32})(c_{21} + c_{23})\\[-2pt]
{\tt 22{A}}&~&(a_{13} + a_{33})(b_{31} + b_{32} + b_{33})(c_{11})\\[-2pt]
{\tt 23{A}}&~&(a_{23})(b_{31} + b_{32} + b_{33})(c_{11} + c_{31} + c_{32})\\ \hline
{\tt 20{B}}&~&(a_{11} + a_{12})(b_{22})(c_{21} + c_{23})\\[-2pt]
{\tt 21{B}}&~&(a_{11})(b_{12} + b_{22} + b_{32})(c_{21} + c_{23})\\[-2pt]
{\tt 22{B}}&~&(a_{13} + a_{33})(b_{31} + b_{32} + b_{33})(c_{31} + c_{32})\\[-2pt]
{\tt 23{B}}&~&(a_{13} + a_{23} + a_{33})(b_{31} + b_{32} + b_{33})(c_{11} + c_{31} + c_{32})
\end{eqnarray*}
\end{minipage}
\caption{Two neighboring schemes 
  with 19 identical summands and 4 different ones.
}
\label{fig:neighbors}
\end{figure}

 \section{Evaluation and Analysis}\label{sec:evaluation}

The methods presented in Section~\ref{sec:methods} enabled us to find several hundreds of solutions individually, but they were particularly
effective when combined.
The first method allows finding schemes that can be quite different compared to the known schemes. However, finding
a scheme using that method may require a few CPU hours as most pairings of type~3 terms cannot be extended to a valid scheme that
satisfies the streamlining constraints. The second method can find schemes that are very similar to known ones with a second. The
neighborhood of known solutions turned out to be limited to a few hundred of new schemes.

\begin{figure}[t]
\centering
\input{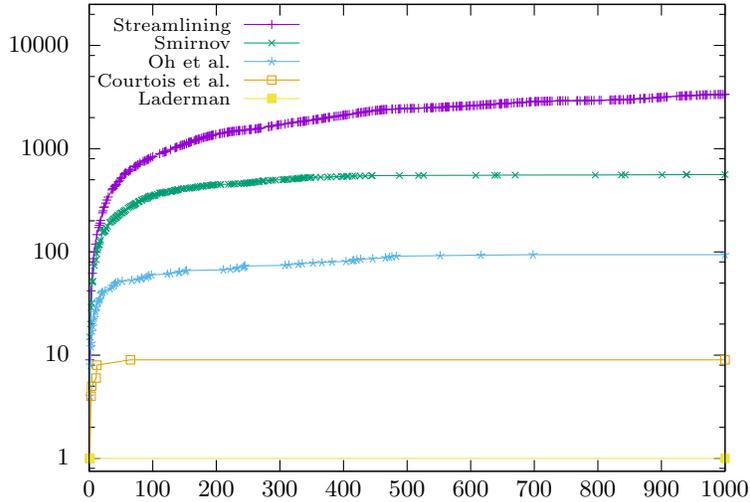}
\caption{The number of different schemes (vertical axis in logscale) found within a period of time (horizontal axis in seconds) during a random walk in the neighborhood of a given scheme.}
\label{fig:results}
\end{figure}

In contrast, some of the schemes found using the first method have a large neighborhood. We approximated the size of the neighborhood
of a scheme using the following experiment: Start with a given scheme $S$ and find a neighboring scheme by randomly fixing 2/3 of the base variables. Once
a neighboring scheme $S'$ is found, find a neighboring scheme of~$S'$, etc. We ran this experiment on a machine with 48 cores of the Lonestar 5 cluster of
Texas Advanced Computing Center. We started each experiment using 48 threads with each thread assigned a different seed.
Figure~\ref{fig:results} shows the number of different schemes (after sorting) found in 1000 seconds when
starting with one of the four known schemes and a scheme that was computed from the streamlining method.
Some of these different schemes are new, while others are equivalent to each other or known ones. We only assure here that they are not identical after sorting the summands.

Observe that the number of different schemes found in 1000 seconds depends a lot on the starting scheme. No different neighboring scheme was found for Laderman's scheme, only 9 different schemes were found for the scheme of Courtois et al., 94 different schemes were found for the scheme of Oh et al., 561 new schemes were found for Smirnov's scheme, and 3359 schemes were found using a randomly selected new scheme obtained with the streamlining method.

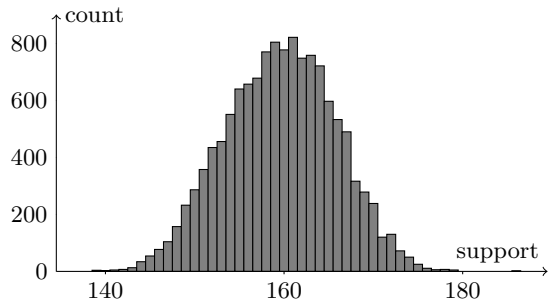
\begin{figure}[t]
\centering
\begin{tikzpicture}[y=.0015in,x=.0275in,xscale=1.7]
\foreach \x/\y in {139/4,140/3,141/5,142/7,143/13,144/34,145/54,146/77,147/104,148/157,149/232,150/286,151/357,152/434,153/455,154/550,155/639,156/656,157/677,158/769,159/803,160/776,161/820,162/747,163/757,164/720,165/596,166/532,167/489,168/316,169/278,170/238,171/120,172/130,173/72,174/50,175/25,176/11,177/6,178/7,179/5,186/2}
\draw[fill=gray] (\x,0) rectangle (\x+1,\y);
\draw[->] (135,0)--(190,0) node[above left] {support};
\draw[->] (135,0)--++(0,900) node[right] {count};
\foreach \x in {140,160,180} \draw (\x.5,0)--++(0,-10) node[below] {\footnotesize\x};
\foreach \y in {0,200,400,...,800} \draw (135,\y)--++(0,-1) node[left] {\footnotesize\y};
\end{tikzpicture}
\caption{Number of non-equivalent schemes found, arranged by support.}
\label{fig:support}
\end{figure}

In view of the large number of solutions we found, it is also interesting to compare them with each other. For example,
if we define the \emph{support} of a solution as the number of base variables set to~$1$, we observe that the support
seems to follow a normal distribution with mean around 160, see Fig.~\ref{fig:support} for a histogram. We can also see
that the Laderman scheme differs in many ways from all the other solutions. It is, for example, the only scheme whose
core consists of four quadruples of type~3 terms. In 89\% of the solutions, the core consists of four pairs of type~3 terms,
about 10\% of the solution have three pairs and one quadrupel, and less than 1\% of the schemes have cores of the form 2-2-2-2-3
or 2-2-2-3-4.

 \section{Challenges}\label{sec:challenges}

The many thousands of new schemes
 that we found may still be just the tip of the iceberg. However, we also observed that the state-of-the-art SAT solving techniques are unable to answer
 several other questions. This section provides four challenges for SAT solvers with increasing difficulty. For each challenge we constructed one or
 more formulas that are available at 
\begin{center}
{\tt \url{https://github.com/marijnheule/matrix-challenges}}. 
\end{center}
The challenges are hard, but they may be
 doable in the coming years.

\paragraph{Challenge 1: Local search without streamlining.}

Our first method combines randomly pairing the type~3 terms with streamlining constraints. The latter was required to
limit the search. We expect that local search solvers can be optimized to efficiently solve the formulas without the streamlining
constraints. This may result in schemes that are significantly different compared to ones we found. We prepared ten
satisfiable formulas with hardcoded pairings of type~3 terms. Five of these formulas can be solved using {\tt yalsat} in a few
minutes. All of these formulas appear hard for CDCL solvers (and many local search solvers).

\paragraph{Challenge 2: Prove unsatisfiability of subproblems.}

We observed that complete SAT solvers performed weakly on our
matrix multiplication instances. It seems therefore unlikely that one could prove any optimality results for
the product of two $3 \times 3$ matrices using SAT solvers in the near future. A more realistic challenge concerns proving
unsatisfiability of some subproblems. We prepared ten formulas with 23 multiplications and hardcoded pairings of type~3 terms.
We expect that these formulas are unsatisfiable.

\paragraph{Challenge 3: Avoiding a type~3 term in a summand.}

All known schemes have the following property: each summand has at least one type~3 term.
We do not know whether there exists a scheme with 23 multiplications such that one of the
summands contains no type~3 term. The challenge problem blocks the existence of a type~3 term
in the last summand and does not have any additional (streamlining) constraints.

\paragraph{Challenge 4: Existence of a scheme with 22 multiplications.}

The main challenge concerns finding a scheme with only 22 multiplications. The hardness
of this challenge strongly depends on whether there exists such a scheme. The repository
contains a plain formula for a scheme with 22 multiplications.

\paragraph*{\bf Acknowledgments.}
The authors acknowledge the Texas Advanced Computing Center at The University of Texas at Austin for providing HPC
resources that have contributed to the research results reported within this paper.

\bibliographystyle{plain}
 \bibliography{all}

\end{document}